\documentstyle{article}

\font\tenrm=cmr10
\font\tenit=cmti10
\font\elevenbf=cmbx10 scaled\magstep 1
\font\elevenrm=cmr10 scaled\magstep 1

\renewenvironment{thebibliography}[1]
 { \tenrm
 \baselineskip=10pt
   \begin{list}{\arabic{enumi}.}
    {\usecounter{enumi} \setlength{\parsep}{0pt}
     \setlength{\itemsep}{3pt} \settowidth{\labelwidth}{#1.}
     \sloppy
    }}{\end{list}}

\begin{document}
\begin{center}{\elevenbf
HYPERNUCLEI AS CHIRAL SOLITONS. \\}
\vglue 0.5cm
{\tenrm V.B.Kopeliovich \\}
{\tenit Institute for Nuclear Research of the Russian Academy of
Sciences,\\ Moscow 117312, Russia\\}
\vglue 0.2cm
\end{center}
{\rightskip=3pc
\leftskip=3pc
\tenrm\baselineskip=10pt
\noindent
The identification of flavored multiskyrmions with the ground states of known 
hypernuclei is successful for several of them, e.g. for isodoublet 
$_\Lambda^4H\,-\, _\Lambda^4He$, isoscalars $_\Lambda^5He$ and $_\Lambda^7Li$. 
In other cases agreement is not so good, but the behaviour of the binding 
energy with increasing baryon number is in qualitative agreement with data.
Charmed or beautiful hypernuclei within this approach are predicted to be 
bound stronger than strange hypernuclei. This conclusion is stable relative
certain variation of poorly known heavy flavor decay constants.
 \vglue 0.2cm}
\baselineskip=14pt
\elevenrm
\section{Introduction}
One of the actual questions of nuclear and elementary particle physics is 
the possibility
of the existence of nuclear matter fragments with unusual properties, 
e.g. with flavor being different from that of $u$ and $d$ quarks. This issue can have interesting consequences in astrophysics and cosmology.The recently observed at Chandra X-ray Observatory stellar objects, RXJ1856 and
3C58, can be interpreted just as the strange quark matter stars. 
Experimental as well as theoretical studies of such nuclear fragments have
been performed first for strangeness, see e.g. \cite{1,2} and references 
therein, and also, to some extent,
for charm and beauty quantum numbers \cite{3}-\cite{6}. Theoretical approaches 
vary from standard nuclear potential models, to the topological soliton models
(Skyrme model and its extensions).
In the latter case, the extension of the original $SU(2)$ model into $SU(3)$ 
configuration space is necessary.
It is known that in $SU(3)$ extensions of the model there are several
different local minima in the configuration space \cite{7}. 
The quantization of configurations near each of these minima is possible, leading to the prediction
of spectrum of quantum states with different flavor quantum numbers.
Here the quantization of $SU(2)$ bound skyrmions embedded into $SU(3)$ is 
considered, following papers \cite{8,9,10}. The physical interpretation of such
quantum states seems to be the simplest in comparison with others, because the
lowest in energy states can be identified with usual nuclei. Previously we
derived in this way some spectrum of "flavored multiskyrmions" regardless of 
their interpretation \cite{10}. Here we make an attempt to identify some of 
these states with known hypernuclei.

The chiral soliton models provide a picture of baryonic systems $(BS)$ outside, 
from large enough distances, based on few fundamental principles and 
ingredients incorporated in
the model lagrangian. The details of baryon-baryon interactions do not enter
the calculations explicitly, although they make influence, of course,
implicitly, via some integral characteristics of $BS$, such as their masses, 
moments of inertia ($\Theta_F,\;\Theta_T$ below), sigma-term ($\Gamma$), etc.
The $SU(2)$ rational map $(RM)$ ansatz \cite{11} which approximates well the 
results of numerical calculations \cite{12} was used as the starting point for
the evaluation of static properties of bound states of skyrmions necessary for
their quantization in the $SU(3)$ configuration space.
The knowledge of the "flavor" moment of inertia and 
the $\Sigma$-term allows us then to estimate the flavor excitation energies 
\cite{8,10}. The masses of the lowest states with strangeness, charm or beauty 
are calculated within the rigid oscillator
version of the bound state approach, and the binding energies of baryonic 
systems with different flavors, $s,\;c$ or $b$, are estimated.
Within the $RM$ approximation, at large enough $B$
the chiral field configuration has the form of the "bubble" with universal
properties of the shell, where the mass and baryon number of the $BS$ are
concentrated. The width of the shell and its average mass density do not depend
on the baryon number \cite{13}. This picture can be acceptable for not
large $B$ ($B=A$ atomic number of the nucleus) say, up to $B\sim 16$, and by this
reason we discuss here the hypernuclei not heavier than hyper-oxygen.

\section{Lagrangian and the mass formula} 
The Lagrangian of the Skyrme model, which in its well known form
depends on parameters $F_{\pi}, \; F_D$ - meson decay constants, the Skyrme
constant $e$, etc., has been presented previously \cite{9,10}, and we give 
here its density for completeness:
$$ {\cal L} = {\cal L}^{(2)} + {\cal L}^{(4)} +{\cal L}^{(6)} + 
{\cal L}^{SB} \eqno (1) $$
with the term of the second order in chiral derivative
$${\cal L}^{(2)}= -{ F_\pi^2\over 16} Tr \, l_\mu l_\mu , $$
the antisymmetric $4-th$ order, or Skyrme term
$${\cal L}^{(4)} = {1\over 32e^2} Tr [l_\mu l_\nu]^2 , $$
the $6-th$ order term
$${\cal L}^{(6)}=c_6 Tr([l_\mu l_\nu][l_\nu l_\gamma][l_\gamma l_\mu]) $$
and the symmetry (chiral and flavor) breaking terms
$${\cal L}^{SB} = {F_\pi^2m_\pi^2 \over 16} Tr\,(U+U^\dagger -2) + $$
$$+\frac{F_D^2m_D^2-F_\pi^2m_\pi^2}{24} Tr(1-\sqrt{3}\lambda_8)(U+U^\dagger-2)+$$
$$ +\frac{F_D^2-F_\pi^2}{48}Tr(1-\sqrt{3}\lambda_8)(Ul_\mu l_\mu +
l_\mu l_\mu^\dagger U^\dagger) \eqno(2) $$
Here the left chiral derivative $l_\mu=\partial_\mu U U^\dagger$, the unitary
matrix $U\in SU(3)$.
In the original Skyrme model the $6$-order term ${\cal L}^{(6)}$, which can be
presented also as a baryon (topological) number density squared, was not 
included, and we shall omit it here as well. Recent calculations of flavor
excitation energies, performed by A.M.Shunderyuk, provide the results which are
close to the results obtained in \cite{10} and in present paper.
The Wess-Zumino term in the action, which can be written as a 5-dimensional 
differential form, plays a very important role in the quantization procedure, 
but it does not contribute to most of static properties of skyrmions, 
see e.g. \cite{8,10}.

The physical values of these constants are: $F_\pi$ = $186$ Mev, $e$ is close
to $e=4$, we take here the value $e=4.12$ \cite{14}. 
The chiral symmetry breaking part of the Lagrangian depends on meson masses,
the pion mass $m_\pi$, and the mass of $K, \, D$ or $B$ meson, we call it 
$m_D$. The flavor symmetry breaking $(FSB)$ part of the Lagrangian is of
the usual form, and was sufficient to describe the mass splittings of the octet
and decuplet of baryons \cite{14}, within the collective coordinates
quantization approach with configuration mixing. It is important that the 
flavor decay 
constant (pseudoscalar decay constant $F_K,\; F_D$ or $F_B$) is different from
the pion decay constant $F_\pi$. Experimentally, $F_K/F_\pi \simeq 1.22$ and 
$F_D/F_\pi\simeq 2.28^{+1.4}_{-1.1} $ \cite{15}. The 
$B$-meson decay constant is not measured yet. In view of this uncertainty,
we take for our estimates two values of $r_c=F_D/F_\pi$, $r_c=1.5$ and $2$, 
and same for $r_b=F_B/F_\pi$, following also to theoretical 
estimates \cite{16}.

We begin our calculations with unitary matrix of chiral fields $U \in SU(2)$, 
as was mentioned above. The classical mass of $SU(2)$ solitons and other static
characteristics necessary for our purposes, in most general case depend on $3$
profile functions: $f, \, \alpha$ and $\beta$.
The general parametrization of $U_0$ for an $SU(2)$ soliton we use here 
is given by $U_0 = c_f+s_f \vec{\tau}\vec{n}$ with $n_z=c_{\alpha}$, 
$n_x=s_{\alpha}
c_{\beta}$, $n_y=s_{\alpha}s_{\beta}$, $s_f=sinf$, $c_f=cosf$, etc. For the 
$RM$ ansatz $f=f(r)$, i.e. the profile depends on one variable, only;
the components of vector $\vec{n}$ are some rational functions of two angular 
variables which define the direction of radius-vector $\vec{r}$ \cite{11}.

The quantization of solitons in $SU(3)$ configuration space was made in the
spirit of the bound state approach to the description of strangeness,
proposed in \cite{17} and used in \cite{18,19}. We use here somewhat simplified
and very transparent variant, the so called rigid oscillator version, proposed
in \cite{8}. The details of the quantization procedure can be
found in \cite{8}-\cite{10}, and we shall not reproduce them here.
We note only, that the $(u,d,c)$ and $(u,d,b)$ $SU(3)$ groups are quite 
analogous to the $(u,d,s)$ one, for the $(u,d,c)$ group a simple redefiniton of 
hypercharge should be made. 

The following mass formula has been obtained for the masses of states with
definite quantum numbers: baryon (topological) number $B$, flavor $F$ 
(strangeness, charm or beauty), isospin $I$ and 
angular momentum $J$ \cite{8,10}:
$$E(B,F,I,J)= M_{B,cl} + |F| \omega_{F,B} +\frac{1}{2\Theta_{T,B}}
\bigl[c_{F,B}T_r(T_r+1)+(1-c_{F,B})I(I+1) + $$
$$ +(\bar{c}_{F,B}-c_{F,B})
I_F(I_F+1)\bigr] +\frac{J(J+1)}{2\Theta_{J,B}},  \eqno(3) $$
$\omega_{F,B}$ or
$\bar{\omega}_{F,B}$ are the 
frequences of flavor (antiflavor) excitations:
$$ \omega_{F,B} = N_cB(\mu_{F,B} -1)/(8\Theta_{F,B}), \qquad
 \bar{\omega}_{F,B} = N_cB(\mu_{F,B} +1)/(8\Theta_{F,B}).\eqno (4)$$
with
$ \mu_{F,B} =[ 1 + 16 \Theta_{F,B}\bigl(\bar{m}_D^2 \Gamma_B+
(F_D^2-F_\pi^2)\tilde{\Gamma}_B\bigr) / (N_cB)^2 ]^{1/2}$,
$N_c$ is the number of colors of the underlying QCD ($N_c=3$ in all numerical
estimates), $\bar{m}^2_D=F_D^2m_D^2/F_\pi^2-m_\pi^2$.
 The terms $\pm N_cB/(8\Theta_{F,B})$ in $(4)$ arise from the Wess-Zumino term
in the action which does not contribute to the masses and momenta of inertia
of skyrmions \cite{17,8}.
In terms of the quark models the difference $\bar{\omega}-\omega =
N_cB/(4\Theta_{F,B})$ is the energy necessary for the production of additional 
$q\bar{q}$ pair. The hyperfine structure constants $c_{F,B}$ and 
$\bar{c}_{F,B}$ are given by \cite{8}
$$c_{F,B} = 1\,- \, \frac{\Theta_{T,B} (\mu_{F,B}-1)}{2\Theta_{F,B}\,
\mu_{F,B}}, \qquad
\bar{c}_{F,B} = 1\,- \, \frac{\Theta_{T,B}(\mu_{F,B}-1)}
{\Theta_{F,B}(\mu_{F,B})^2}.\eqno(5) $$
Evidently, when $\mu \to \infty$, $\bar{c} \to 1$.
The contributions of the order of $1/\Theta \sim N_c^{-1}$ which depend
originally on angular velocities of rotations in isospace and usual space are 
taken into account in $(3)$. This expression was obtained
by means of quantization of the oscillator-type hamiltonian describing the
motion of the $SU(2)$ skyrmion in the $SU(3)$ collective coordinates space.
The classical mass $M_{cl}\sim N_c$, and the energies $\omega_F \sim N_c^0 =1$.
The motion into "flavor" direction, $s, c$ or $b$ is described by
the amplitude $D$ \cite{8,10} which is small for the lowest quantum states 
(lowest $|F|$):
$D \sim \bigl[16\Gamma_B\Theta_{F,B}\bar{m}_D^2 + N_c^2B^2 
\bigr]^{-1/4}(2|F|+1)^{1/2}$.
So, the amplitude $D$ decreases with increasing mass $m_D$ like $1/\sqrt{m_D}$,
as well as with increasing number of colors $N_c$, and the method works for 
any value of mass $m_D$, also for charm and beauty quantum numbers.

In $(3)$  $I$ is the isospin of the multiplet with flavor $F$, $T_r=p/2$ is the so 
called "right" isospin - the isospin of the not-flavored component of the
$SU(3)$ multiplet under consideration with $(p,q)$ - the numbers of upper and
lower indices in the spinor which describes it. $I_F$ is the isospin carried
by flavored mesons which are bound by $SU(2)$ skyrmion, $\vec{I}= \vec{T_r}
+\vec{I_F}$. Evidently, $ I_F \leq |F|/2$.
In the rigid oscillator model the states predicted do not correspond to
the definite $SU(3)$ or $SU(4)$ representations. How they can be ascribed to
them was shown in \cite{8,10}.  
For example, the state with $B=1$, $|F|=1$, $I=0$ should belong to the 
octet of $(u,d,s)$, or $(u,d,c)$, $SU(3)$ group, if $N_c=3$. Here we consider
the quantized states of $BS$ which belong to the lowest
possible $SU(3)$ irreps $(p,q)$, $p+2q=3B$: $p=0, \; q=3B/2$ for
even $B$, and $p=1, \; q=(3B-1)/2$ for odd $B$. For $B=3,\, 5$ and $7$ they 
are $\bar{35}, \, \bar{80}$ and $\bar{143}$-plets,
for $B=4, \, 6$ and $8$ - $\bar{28}$,  $\bar{55}$ and $\bar{91}$-plets, etc.
For even $B, \; T_r=0$, for odd $B$, $T_r=1/2$ for the lowest 
$SU(3)$ irreps (see Fig. 1).

\vglue 1.5cm


\def\br{\mbox{\boldmath $r$}}
\def\bm{\mbox{\boldmath $m$}}

\setlength{\unitlength}{1.3cm}
\begin{flushleft}

\begin{picture}(12,6)
\put(3,1.5){\vector(1,0){2.5}}
\put(3,1.5){\vector(0,1){3.8}}
\put(2.8,5.2){$Y$}
\put(5.5,1.2){$I_3$}
\put(1,5.7){ a) $Odd \;B\,,\; J=1/2$}
\put(2.3,4.2){$^3 H$}
\put(3.3,4.2){$^3 He$}
\put(2.8,3.3){$^3_\Lambda H$}
\put(2.5,4){\circle {0.15}}
\put(3.5,4){\circle {0.15}}
\put(2,3){\circle {0.15}}
\put(3,3){\circle {0.15}}
\put(3,3){\circle {0.27}}
\put(4,3){\circle {0.15}}
\put(1.5,2){\circle {0.15}}
\put(2.5,2){\circle {0.15}}
\put(3.5,2){\circle {0.15}}
\put(4.5,2){\circle {0.15}}



\put(9,1.5){\vector(1,0){2.3}}
\put(9,1.5){\vector(0,1){3.8}}
\put(8.8,5.2){$Y$}
\put(11,1.2){$I_3$}
\put(7,5.7){ b) $Even \;B\,,\; J=0$}
\put(8.8,4.2){$^4 He $}
\put(8.3,3.3){$^4_\Lambda H$}
\put(9.3,3.3){$^4_\Lambda He$}

\put(9,4){\circle {0.15}}

\put(8.5,3){\circle {0.15}}
\put(8.5,3){\circle {0.27}}
\put(9.5,3){\circle {0.15}}
\put(9.5,3){\circle {0.27}}

\put(8,2){\circle {0.15}}
\put(9,2){\circle {0.15}}
\put(10,2){\circle {0.15}}

\end{picture}
\end{flushleft}

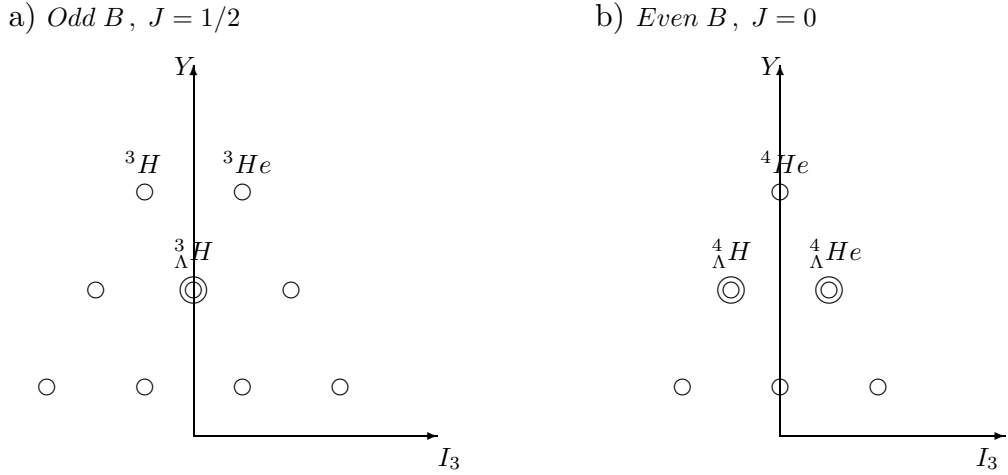
{\bf Fig. 1}. (a) The location of the isoscalar state (shown by double circle)
with odd $B$-number and $|F|=1$ in the upper part of the $(I_3 -Y)$ diagram.
(b) The same for isodoublet states with even $B$. The case of light hypernuclei
$_\Lambda H$ and $_\Lambda He$ is presented as an example. The lower parts of 
diagrams with $Y \leq B-3$ are not shown here.
\\

The flavor moment of inertia which enters the above formulas
 \cite{8,10,17}
for arbitrary $SU(2)$ skyrmions is given by \cite{10}:
$$ \Theta_F = {1 \over 8} \int (1-c_f)\biggl\{ F_D^2 + {1 \over e^2}
 \bigl[ (\vec{\partial}f)^2 +s_f^2(\vec{\partial}n_i)^2 \bigr] \biggr\} 
d^3\vec{r}. \eqno (6) $$
$(\vec{\partial}n_i)^2 =(\vec{\partial}\alpha)^2 +s_\alpha^2
(\vec{\partial}\beta)^2 $.
It is simply connected with $\Theta_F^{(0)}$ for the flavor symmetric case:
$ \Theta_F=\Theta_F^{(0)}+(F_D^2/F_\pi^2-1) \Gamma/4,$
$\Gamma$ is defined in $(7)$ below.
The flavor inertia increases with $B$ almost proportionally to $B$. 
The isotopic moments of inertia are the diagonal components of the 
corresponding tensor of inertia, in our case this tensor of 
inertia is close to unit matrix multiplied by $\Theta_T$.

The quantities $\Gamma$ (or $\Sigma$-term), which defines
the contribution of the 
mass term to the classical mass of solitons, and $\tilde{\Gamma}$ 
in $\mu_{F,B}$ are given by:
$$ \Gamma = \frac{F_{\pi}^2}{2} \int (1-c_f) d^3\vec{r}, \qquad
  \tilde{\Gamma}={1 \over 4} \int c_f
 \bigl[ (\vec{\partial}f)^2 +s_f^2(\vec{\partial}n_i)^2
 \bigr] d^3\vec{r}. \eqno (7)$$
For the $RM$ ansatz the formulas $(6),\;(7)$ can be slightly modified 
\cite{10}, but in such general form they look simple enough already.
The masses of solitons have been calculated in \cite{12} and \cite{10}, 
moments of inertia, $\Gamma$ and $\tilde{\Gamma}$ were calculated in \cite{10} 
for several values of $B$, the missing quantities are calculated here. 
The contribution to the $\mu_{F,B}$ proportional to $\tilde{\Gamma}_B$ is 
suppressed in comparison with the term $\sim \Gamma$ by the small factor 
$\sim F_D^2/m_D^2$, and is more important for strangeness.
\section{Strange and beautiful hypernuclei}
It is convenient to calculate the energy difference between the state with 
flavor $F$ belonging to the
$(p,q)$ irrep, and the ground state with $F=0$ and the same $B,\;J$
and $(p,q)$ \cite{10}:
$$ \Delta E_{B,F} = |F| \omega_{F,B} + \frac{\mu_{F,B}-1}{4\mu_{F,B}
\Theta_{F,B}}[I(I+1)-T_r(T_r+1)]
 + \frac{(\mu_{F,B}-1)(\mu_{F,B}-2)}{4\mu_{F,B}^2 \Theta_{F,B}} 
I_F(I_F+1), \eqno (8) $$
It was used in deriving $(3)$ and $(8)$ that the so called "interference"
moment of inertia which makes a contribution to the lagrangian proportional 
to the product of angular velocities of rotation in the isotopic and ordinary 
$3D$ space, is negligible compared to the isotopic and orbital tensors of 
inertia \cite{20} for all multiskyrmions except those with $B=1,2$.
Note also, that $(8)$ does not depend on $\Theta_T$, only on $\Theta_F$, when
the formulas for hyperfine splitting constants are used.

For the state with isospin $I=0$ and unit flavor number, $|F|=1$, the binding 
energy difference in comparison with the ground state of the nucleus with the 
same $B,\; (p,q)$ and $|F|=0$, is
$$\Delta \epsilon_{B,F}= \omega_{F,1}-\omega_{F,B} - \frac{3(\mu_
{F,1}-1)}{8\mu_{F,1}^2\Theta_{F,1}}
+ \frac{3(\mu_{F,B}-1)}{8\mu_{F,B}^2\Theta_{F,B}} \eqno (9) $$
Such states can exist for odd $B$, with $I_F=T_r=1/2$, see Fig.1a.
In the case of antiflavor excitations we have similar formulas, with the
substitution $ \mu \to -\mu $. 
\begin{center}
\begin{tabular}{|l|l|l|l|l||l|l|l||l|l|l|}
\hline
 $_\Lambda A$  &$\omega_s$& $\Delta \epsilon_s $ & $\epsilon^{tot}_s$&
 $\epsilon^{tot}_{exp,s}$
 &$\omega_b^{r_b=1.5}$&$\Delta \epsilon_b$ &$\epsilon^{tot}_b$
 &$\omega_b^{r_b=2}$&$\Delta \epsilon_b$ &$\epsilon^{tot}_b$ \\
\hline
$1$ &$306$&---&---&---&$4501$&---&--- &$4805$&---&--- \\
\hline
$^3_\Lambda H$&$ 289$ &$-3$  &$\;5$&$2.35$
&$4424$&$75$&$83$ &$4751$&$53$&$61$ \\
\hline
$^5_\Lambda He$&$ 287$&$- 6$&$33 $&$31.4$
&$4422$&$76$&$103$&$4749$&$54$&$81$ \\
\hline
$^7_\Lambda Li$&$282 $ &$-3$&$29$&$37.6$
&$4429$&$81$&$119$ &$4744$&$59$&$97$ \\
\hline
$^9_\Lambda Be$&$291 $&$-13$&$40$&$63.2$
&$4459$&$40$&$97$ &$4773$&$31$&$88$ \\
\hline
$^{11}_\Lambda B$&$294 $&$-16$&$59$&--- 
&$4478$&$21$&$96$ &$4786$&$18$&$93$ \\
\hline
$^{13}_\Lambda C$&$295$&$-18$&$78$&$104$
&$4488$&$10$&$106$ &$4793$&$11$&$107$ \\
\hline
$^{15}_\Lambda N$&$300$&$-23$&$91$&$118$
&$4515$&$-17$&$97$ &$4810$&$-7$&$108$ \\
\hline
\end{tabular}
\end{center}
\vspace{1mm}

{\rightskip=3pc
\leftskip=3pc
\tenrm\baselineskip=12pt 
\noindent
{\bf Table 1.}  The collective motion contributions to the binding energies 
of the isoscalar hypernuclei with unit 
flavor, strangeness or beauty, $S=-1$ or $b=-1$, in $Mev$. 
$\omega_s$ and $\omega_b$ are the strangeness and beauty excitation energies, 
$\Delta \epsilon_{s,b}$, in $Mev$, are the changes of binding energies of 
lowest $BS$ with flavor 
$s$ or $b$, $|F|=1$, in comparison with usual $(u,d)$ nuclei with the same 
B-number. $\epsilon^{tot}$ is the total binding energy of the hypernucleus.
Experimental values $\epsilon^{tot}_{exp}$ are taken from \cite{1,2}. 
The energies $\omega$ for $B=1$
are given for comparison. For beauty the first 3 columns correspond to
$r_b=F_B/F_\pi =1.5$, and the last 3 - to $r_b = 2$. 
\vglue 0.1cm}

\begin{center}
\begin{tabular}{|l|l|l|l|l||l|l|l||l|l|l|}
\hline
 $_\Lambda A$  &$\omega_s$ &$\Delta\epsilon_s$ & $\epsilon^{tot}_s$&
 $\epsilon^{tot}_{exp}$
 &$\omega_b^{r_b=1.5}$&$\Delta \epsilon_b$ &$ \epsilon^{tot}_b$
 &$\omega_b^{r_b=2}$&$\Delta \epsilon_b$ &$ \epsilon^{tot}_b$ \\
\hline
$^4_\Lambda H - ^4_\Lambda He$&$283$&$-23$&$5.3$&$10.52;10.11$&
$4402$&$71$&$99$ &$4735$&$52$&$80$ \\
\hline
$^6_\Lambda He - ^6_\Lambda Li$&$287$&$-22$&$10.3$&$31.7;\; \;30.8$&
$4430$&$52$ &$84$&$4752$&$40$ &$72$\\
\hline
$^8_\Lambda Li - ^8_\Lambda Be$&$288 $&$-20$&$36.5$&$46.05; \;44.4$&
$4443$&$43$ &$99$&$4765$&$33$&$89$\\
\hline
$^{10}_\Lambda Be -^{10}_\Lambda B$&$292$&$-23$&$42$&$67.3; \;\;65.4$&
$4465$&$24$&$89$&$4778$&$20$&$85$\\
\hline
$^{12}_\Lambda B - ^{12}_\Lambda C$&$294$&$-24$&$67$&$87.6;\;\; 84.2$&
$4481$&$10$&$102$&$4788$&$11$&$103$\\
\hline
$^{14}_\Lambda C -^{14}_\Lambda N$&$299$&$-28$&$77$&$109.3 ; 106.3$&
$4506$&$-14$ &$91$&$4805$&$-5$ &$100$\\
\hline
$^{16}_\Lambda N -^{16}_\Lambda O$&$301$&$-30$&$97$& ---&
$4521$&$-28$ &$100$&$4815$&$ -14$ &$114$\\
\hline
\end{tabular}
\end{center}
\vspace{1mm}

{\rightskip=3pt
\leftskip=3pt
\tenrm\baselineskip=12pt
{\bf Table 2.} The binding energies of the 
isodoublets of hypernuclei with unit flavor, $S=-1$ or $b=-1$ in $Mev$.
Other notations and peculiarities as in {\bf Table 1.}
\vglue 0.1cm}

For the states with maximal isospin $I=T_r+|F|/2$ 
the energy difference can be simplified to \cite{10}:
$$\Delta E_{B,F}=|F|\biggl[\omega_{F,B}+ T_r \frac{\mu_{F,B}-1}
{4\mu_{F,B}\Theta_{F,B}}
+\frac{(|F|+2)}{8\Theta_{F,B}} \frac{(\mu_{F,B}-1)^2}{\mu_{F,B}^2} \biggr].
\eqno (10) $$
The case of isodoublets, even $B$, is described by $(8)$ with $T_r=0$, see
{\bf Table 2} and Fig.1b.
It follows from $(10)$ that when a nucleon is replaced by a
flavored hyperon in $BS$ the binding energy of the system with $|F|=1, \;T_r=0$ 
changes by
$$\Delta \epsilon_{B,F}=\omega_{F,1}-\omega_{F,B} - \frac{3(\mu_
{F,1}-1)}{8\mu_{F,1}^2\Theta_{F,1}}
- \frac{3(\mu_{F,B}-1)^2}{8\mu_{F,B}^2\Theta_{F,B}}
\eqno (11) $$
For strangeness Eq. $(11)$ is negative indicating that stranglets should 
have binding energies smaller than those of nuclei with the same B-number.

To obtain the values of total binding energy of hypernuclei shown in
{\bf Tables}, we add the calculated difference of binding energies given by
$(9)$ or $(11)$ to the known value of binding energy of usual $(u,d)$ nucleus.
E.g., for $B=3$ it is the average of binding energies of $^3 H$ and $^3He$, 
for $B=4$ it is the binding energy of $^4 He$ $(5.3\;Mev = (28.3-23)\;Mev)$, 
etc., see Fig.1.
A special care should be taken about spin of the nucleus. For $^3_\Lambda H$
and $^3H$, $^4_\Lambda He$ and $^4 He$, $^6_\Lambda Li$ and $^6 Li$, 
$^{13}_\Lambda C$ and $^{13}C$, and in few other cases the spins of the ground
states of hypernucleus and nucleus coincide. For $^5_\Lambda He \; (J=1/2)$
and $^5He\; (J=3/2)$, $^9_\Lambda Be\;(J=1/2)$ and $^9Be\; (J=3/2)$,
$^{12}_\Lambda C\;(J=1)$ and $^{12}C\; (J=0)$ and in some other cases the 
difference in the rotation energies $E_J=J(J+1)/(2\Theta_J)$ should be taken 
into account. E.g., for $^7_\Lambda Li$ this difference decreases the 
theoretical value 
of binding energy by about $7\;Mev$, we have $29\; Mev$ instead of $36$ Mev.
In those cases when the spin of hypernucleus is not known, this correction
was not included in {\bf Tables 1,2}. 
Beginning with $B \sim 10$, the correction to the energy of quantized states 
due to nonzero angular
momentum is small and decreases with increasing $B$ since the corresponding
moment of inertia increases proportionally to $\sim B^2$.

Since $\Theta_{F,B}$ increases with increasing $B$ and
$F_D\;(m_D)$ this leads to the increase of binding with increasing $B$ and 
mass of the "flavor", in agreement with \cite{9,10}. For beauty (and charm, see
below) Eq. $(11)$ is positive for $3 \leq B \leq 12$. As it follows from 
{\bf Tables 1,2}, our method underestimates the binding energy of strangeness 
in nuclei, beginning with $B=A \sim 9$. It means that the other sources of binding should
be taken into account, besides the collective motion of $BS$ in the $SU(3)$
configuration space.
\section{Charmed hypernuclei}

In this section the binding energies of charmed hypernuclei are presented for
two values of the charm decay constant which correspond to the ratio
$r_c=F_D/F_\pi=1.5$ and $r_c=2$. Although the measurement of this constant 
has been performed in \cite{15}, in view of its big uncertainty variation of
this constant in some interval seems to be reasonable.
As it follows from {\bf Table 3}, the predicted binding energies of charmed
hypernuclei differ not essentially for the values $r_c=1.5$ and $r_c=2$, this
difference being invreasing with increasing atomic number.
For light hypernuclei this difference is considerably smaller than for beauty
quantum number (see Section 3).

For the charm, the repulsive Coulomb interaction is greater than for ordinary 
nuclei with the same atomic number. Moreover, since the charmed nucleus has
somewhat smaller dimensions than the ordinary nuclei - the effect which has not
been taken into account by present consideration - this repulsion can decrease
the binding energies for charm by several $Mev$. This, however, does not change 
our qualitative conclusions. For $B=A=5$ and $13$ our results shown in {\bf 
Table 3} agree, within $15-20\;Mev$ with early result by Dover and Kahana
\cite{4} where binding of the charm by several nuclei has been studied within
potential approach. In general, we can speak about qualitative agreement with 
results of such approach for $B \sim 5 \;-\; 10$ \cite{5,6}
(the results of the potential approach have been reviewed in \cite{6}).

\begin{center}
\begin{tabular}{|l|l|l|l||l|l|l|}
\hline
 $_\Lambda A$ &$\omega_c^{r_c=1.5}$&$\Delta \epsilon_c$& $\epsilon^{tot}_c$&
 $\omega_c^{r_c=2}$ &$\Delta \epsilon_c$& $\epsilon^{tot}_c$  \\
\hline
$1$&$1535$&$-$&$-$&$1673$&$-$&$-$ \\
\hline
$^3_\Lambda He$&$1504$&$27$&$35$&$1647$&$24$&$32$\\
\hline
$^5_\Lambda Li$&$1505$&$25$&$52$&$1646$&$25$&$52$ \\
\hline
$^7_\Lambda Be$&$1497$&$32$&$70$&$1641$&$30$&$68$ \\
\hline
$^9_\Lambda B$&$1518$&$11$&$68$&$1654$&$17$&$74$\\
\hline
$^{11}_\Lambda C$&$1525$&$4$&$79$&$1658$&$13$&$87$\\
\hline
$^{13}_\Lambda N$&$1529$&$\;0$&$96$&$1660$&$10$&$106$ \\
\hline
$^{15}_\Lambda O$&$1540$&$-11$&$103$&$1668$&$3$&$117$ \\
\hline
\end{tabular}

{\bf Table 3a.}

\vspace{3mm}
\begin{tabular}{|l|l|l|l||l|l|l|}
\hline
 $_\Lambda A$ &$\omega_c^{r_c=1.5}$&$\Delta \epsilon_c$&$\epsilon^{tot}_c$ 
 &$\omega_c^{r_c=2}$&$\Delta \epsilon_c$&$\epsilon^{tot}_c$ \\
\hline
$^4_\Lambda He - ^4_\Lambda Li$&$1493$&$12$&$40$&$1639$&$16$&$44$ \\
\hline
$^6_\Lambda Li - ^6_\Lambda Be$&$1504$&$\;9$&$41$&$1646$&$14$&$46$\\
\hline
$^8_\Lambda Be - ^8_\Lambda B$&$1510$&$\;7$&$63$&$1648$&$15$&$71$\\
\hline
$^{10}_\Lambda B -^{10}_\Lambda C$&$1520$&$\;0$&$65$&$1655$&$10$&$75$\\
\hline
$^{12}_\Lambda C - ^{12}_\Lambda N$&$1526$&$\,-4$&$88$&$1659$&$\;7$&$99$\\
\hline
$^{14}_\Lambda N -^{14}_\Lambda O$&$1536$&$-14$&$91$&$1666$&$\;1$&$106$\\
\hline
$^{16}_\Lambda O -^{16}_\Lambda F$&$1543$&$-19$&$109$&$1670$&$-2$&$126$\\
\hline
\end{tabular}

{\bf Table 3b.}
\end{center}
\vspace{1mm}

{\rightskip=3pt
\leftskip=3pt
\tenrm\baselineskip=12pt
{\bf Table 3.} The binding energies of the charmed hypernuclei, isoscalars in
{\bf Table 3a} 
and isodoublets in {\bf Table 3b}, with unit charm, $c=1$ in $Mev$. 
$\Delta \epsilon_c$, 
in $Mev$, and $\epsilon^{tot}$ is the same as in {\bf Tables 1,2}, for the
charm quantum number. The results are shown for two values of charm decay
constant, corresponding to $r_c=1.5$ and $r_c=2$. The chemical symbol is 
ascribed to the nucleus according to its total electric charge. 
\vglue 0.1cm}

As in the $B=1$ case, the absolute values of masses of
multiskyrmions are controlled by the poorly
known loop corrections to the classic masses, or the Casimir energy \cite{21}. 
 And as was done for the $B=2$ states, 
 the renormalization procedure is necessary to obtain physically reasonable
values of the masses of multibaryons. 
This generates an uncertainty of about few tens of $Mev$, 
as the binding energy of the deuteron is $30 \; Mev$ instead of the measured 
value $2.225 \; Mev$, so $\sim 30 \; Mev$ characterises the uncertainty of our 
approach \cite{10}. This uncertainty is cancelled mainly in the differences of
binding energies $\Delta \epsilon$ shown in {\bf Tables 1-3}.
\section{Comments and conclusions} 

The version of the bound state soliton model proposed in \cite{8} and
modified in \cite{9,10} for the flavor symmetry breaking case 
$(F_D \; > \; F_\pi)$, allows to 
calculate the binding energy differences of ground states of flavored and
unflavored nuclei, and combined with few phenomenological arguments it is
very successful in some cases of light hypernuclei, e.g. isoscalars
$_\Lambda ^5He$ and $_\Lambda ^7Li$. In other cases the accuracy of
binding energies description is at the level of $10\;-\;30\;Mev$, expected 
for the whole method, which takes into account the collective motion of the
baryonic systems, only. There is also general qualitative agreement with data 
in the behaviour of binding energy with increasing atomic number.It should be stressed that it is, probably, one of interesting examples when 
field theoretical model provides the results which can be directly compared 
with observation data, and can be considered as an additional argument in favor
of applicability of the chiral soliton approach to the description of realistic
properties of nuclei. For the charm and beauty quantum numbers the results 
only slightly depend on the poorly known values of the decay constants $F_D$ 
or $F_B$.

The tendency of decrease of binding energies with increasing 
$B$-number, beginning with $B\sim 10$, is connected with the fact that the 
rational map approximation, leading to the one-shell bubble structure of the 
classical
configuration \cite{11,12,13}, is not good for such values of $B$. At large 
values of the $FSB$ mass we have approximately $\omega_F \simeq m_D
\sqrt{\Gamma/\Theta_F} \;\;F_D/(2F_\pi)$. For $RM$ configurations at large $B$ 
the sigma-term $\Gamma$ grows faster than the inertia $\Theta_F$, because the
contribution of the volume occupied by the chiral field configuration, is more
important for $\Gamma$ \cite{13}. For larger $B=A$, beginning with several tens,
the configurations of the type of skyrmion crystals seem to be more realistic
than $RM$ type configurations.

The variation of the only model parameter, Skyrme constant $e$, makes small 
influence on the results presented here, negligible for charm or beauty quantum
numbers. The quantities $\Gamma$ and inertia $\Theta_F$ both scale like
$1/(F_\pi e^3)$, if the pion mass term in the minimized classical mass is 
omitted, 
therefore, the flavor excitation energies given by $(4)$, depending on their
ratio at large values of $m_D$, are scale invariant. The inclusion of the pion 
mass term slightly changes this conclusion, more for strangeness.

The hypernuclei with $|F| \geq 2$ can be studied using similar methods 
\cite{10}, and it will be done elsewhere.
Consideration of the hypernuclei with "mixed" flavors is possible in
principle, but is technically more involved. For example, the isodoublet 
$^3_{s,c}H$- $^3_{s,c}He$
consisting of $(n, \Lambda, \Lambda_c)$ and $(p, \Lambda,\Lambda_c)$ is 
expected.

There is rough agreement of our results 
with the results of \cite{19,20} where the flavor excitation frequences 
had been calculated within another version of the the bound state approach,
and using the collective coordinates quantization method, for strangeness. 
Some difference in details takes place, however, and it would be of interest to
reproduce our results within other variants of chiral soliton model. The model 
we used overestimates the strangeness excitation energies, but is more reliable
for differences of energies which enter $(9),(11)$, and for charm and beauty
quantum numbers.
Further theoretical studies and experimental searches for the baryonic systems 
with flavor different from $u$ and $d$ could shed more light on the dynamics of heavy flavors in baryonic systems.
Results of this work have been presented at the workshops on Physics on Japan 
Hadron Facility NP01 (December 2001), NP02 (September 2002) and Particle and
Nuclear Physics International Conference, PANIC02 (October 2002). I'm indebted 
to A.M.Shunderyuk for checking numerical calculations and to V.Andrianov, 
A.Gal, T.Nagae for discussions and remarks. 
The work has been supported by the Russian Foundation for Basic Research,
grant 01-02-16615.\\
\vglue 0.2cm

\end{document}